\documentclass[sigconf, anonymous=false]{acmart} 
\setcopyright{none}
\renewcommand\footnotetextcopyrightpermission[1]{}

\setcopyright{acmlicensed}
\copyrightyear{2024}
\acmYear{2024}
\acmDOI{} 
 \acmBooktitle{CONSEQUENCES Workshop at RecSys ’24, October 14, 2024, Bari. ACM, New York, NY, USA,}
\acmISBN{}
\usepackage{multirow}

\acmConference[CONSEQUENCES'24@RecSys]{CONSEQUENCES 2024}{October 14th, 2024}{Bari, Italy}

\begin{document}

\title{Measuring Recency Bias In Sequential Recommendation Systems}

\author{Jeonglyul Oh}
\email{jamesoh0813@bdai.snu.ac.kr}
\affiliation{%
    \institution{Seoul National University}
  \city{Seoul}
  \country{Republic of Korea }
}
\author{Sungzoon Cho}
\authornote{Corresponding author}
\email{zoon@snu.ac.kr}
\affiliation{%
    \institution{Seoul National University}
  \city{Seoul}
  \country{Republic of Korea}
}

\renewcommand{\shortauthors}{Oh et al.}

\begin{abstract}
Recency bias in a sequential recommendation system refers to the overly high emphasis placed on recent items within a user session. This bias can diminish the serendipity of recommendations and hinder the system's ability to capture users' long-term interests, leading to user disengagement. We propose a simple yet effective novel metric specifically designed to quantify recency bias. Our findings also demonstrate that high recency bias measured in our proposed metric adversely impacts recommendation performance too, and mitigating it results in improved recommendation performances across all models evaluated in our experiments, thus highlighting the importance of measuring recency bias. 
\end{abstract}
\maketitle

\section{Introduction} 
Recommendation systems have been a major field in artificial intelligence, with extensive work dedicated to advancing its capabilities. Within this domain, sequential recommendation has emerged as an area of focus, aiming to predict the next item a user is most likely to prefer based on their interaction history. \cite{recency_dropout} introduces the concept of \textit{recency bias}, defining it as the tendency of a recurrent neural network(RNN)-based sequential recommendation model to overly focus on recent user interactions, thereby failing to capture long-term dependencies. Including this and our new perspective, we highlight two critical issues of high recency bias:
\begin{itemize}
    \item A recommendation system with high recency bias seriously lacks serendipity, which is one of the major requirements a fine recommendation system ought to satisfy, as it repeatedly suggests items similar to only those the user recently interacted with. This monotony can quickly bore out user, and further lead to user disengagement.
    \item High recency bias denotes the system’s incapability to fully leverage the entire user session to infer preferences, as noted in \cite{recency_dropout}.  
\end{itemize}
Though \cite{recency_dropout} suggests their way of measuring recency bias, this measure is defined only for reinforcement learning-based models.
Therefore we seek to suggest a more universal metric that is applicable to any sequential recommendation model. 
The contributions of this paper are:
\begin{itemize}
    \item We introduce a simple yet effective novel metric to measure recency bias in sequential recommendation models.
    \item We provide extensive experimental evidence demonstrating that high recency bias results in a decline in the performance of sequential recommendation models.
\end{itemize}
\section{Measuring Recency Bias}\label{section_2}
In this section, we introduce a novel metric designed to quantify recency bias in a sequential recommendation system. Consider the task of sequential recommendation where a user session is represented as \( s = \{i_1, i_2, \dots, i_{n}\} \), and the system recommends an item \( i_k \).
Which item will \( i_k \) be, when recency bias is the greatest? That is, the recommendation system focuses highly on only recent items, and the monotonous recommendation results cause the user to get bored. 
We conclude that recency bias is greatest when \( i_k  = i_n\), the very last item in the session. 
This situation arouses two problems:
first, recommending exactly the last item denotes repetitive and uninteresting recommendation, reducing user satisfaction; second, from a modeling perspective, it suggests that the system has failed to effectively capture the user's broader preferences, instead defaulting to merely recommending the most recent item itself.
To measure the frequency of such a case, we propose a metric that measures the Hit Rate of the Last Item in a session, denoted as HRLI. In doing so, we can now measure recency bias as well, as high HRLI directly correlates with the aforementioned issues of high recency bias. 
By replacing the ground-truth(GT) item with the last item in calculating Hit Rate@K 
of evaluation metric,
the measured target is switched from recommendation performance to recency bias. 
Relevance scores for ranking are calculated on the entire item set, as proposed by \cite{full_eval}. 
Formally, HRLI is defined as: 
\begin{equation}
    HRLI@K = \frac{N(\textit{Hit of Last Item in Top-K})}{N(\textit{Eval set})} ,
\end{equation}
and HRLI is proportional to the degree of recency bias of a sequential recommendation model and is a model-agnostic metric applicable to any architecture. 
\begin{table*}[ht!] 
    \resizebox{\textwidth}{!} {        \begin{tabular}{@{}c|c|cccccc|c|cccccc }
    \toprule
         Dataset&  Metric &  GRU4Rec & SASRec  &STAMP & LightSANs   &  CORE  &FEARec &  Dataset&GRU4Rec & SASRec  &STAMP 
& LightSANs   & CORE  & FEARec \\
         \midrule
         
         \multirow{13}{*}{Beauty}&  HRLI@10 &  0.2204&  0.9915&0.2682
& 0.9809&    0.6564&0.9931&  \multirow{13}{*}{Sports}& 0.0908& 0.9929&0.1516&  0.9915& 0.5311 & 0.9929\\ 
                    \cmidrule{2-8}
                    \cmidrule{10-15}
                    &  NDCG@5 &   0.0245&   0.0327&0.0335&   0.0349&  0.0141& 0.0336&  &0.0116& 0.0172&0.0176
& 0.0187& 0.0063 & 0.0182\\ 
                    &  NDCG*@5&  0.0255&    0.0446&0.0354
&  0.0460&  0.0183&  0.0458&  & 0.0119&  0.0246&0.0179
&  0.0263& 0.0083& 0.0253\\
 & \textbf{Improv}.& \textbf{+4.08\%}& \textbf{+36.39\%}&\textbf{+5.67\%}& \textbf{+31.81\%}& \textbf{+29.79\%}&\textbf{ +36.31\%}& & \textbf{+2.59\%}&\textbf{+43.02\%}&\textbf{+1.70\%}& \textbf{+29.10\%}&\textbf{+31.74\%}& \textbf{+39.01\%} \\ 
                    \cmidrule{2-8}
                    \cmidrule{10-15}
                    &  NDCG@10 &    0.0314&   0.0420&0.0393
&  0.0448&    0.0256& 0.0431&  &0.0155& 0.0224&0.0213
& 0.0244& 0.0121 & 0.0234\\ 
                    &  NDCG*@10& 0.0324&  0.0535&0.0417
& 0.0547&  0.0284& 0.0544&  &0.0158& 0.0296&0.0216
& 0.0315& 0.0144& 0.0307\\
 & \textbf{Improv.}& \textbf{+3.18\%}& \textbf{+27.38\%}&\textbf{+6.11\%}&\textbf{+22.10\%}& \textbf{+10.94\%}& \textbf{+26.22\%}& &\textbf{+1.94\%}& \textbf{+32.14\%}&\textbf{+1.41\%}& \textbf{+12.65\%}& \textbf{+19.00\%}& \textbf{+31.20\%} \\ 
                    \cmidrule{2-8}
                    \cmidrule{10-15}
                    &  Hit@5 &    0.0368&  0.0566&0.0472
&   0.0588&   0.0279&  0.0566&  &0.0182& 0.0320&0.0247
& 0.0340& 0.0135 & 0.0335\\ 
                    &  Hit*@5&  0.0379&    0.0635&0.0492
&  0.0669&  0.0336& 0.0647&  &0.0185& 0.0356&0.0251
& 0.0383& 0.0164& 0.0367\\
 & \textbf{Improv.}&\textbf{+2.99\%}&\textbf{+12.19\%}&\textbf{+4.24\%}& \textbf{+13.78\%}& \textbf{+20.43\%}&\textbf{ +14.31\%}& & \textbf{+1.65\%}& \textbf{+11.25\%}&\textbf{+1.62\%}& \textbf{+4.23\%}&\textbf{+21.48\%}&\textbf{+9.55\%}\\ 
                    \cmidrule{2-8}
                    \cmidrule{10-15}
                    &  Hit@10 &  0.0580&  0.0854&0.0655
& 0.0894&  0.0635& 0.0860&  &0.0305& 0.0485&0.0361
& 0.0520& 0.0315 & 0.0498\\ 
                    &  Hit*@10& 0.0593&  0.0914&0.0688& 0.0936&  0.0649& 0.0916&  &0.0306& 0.0511&0.0367&  0.0542& 0.0357& 0.0535\\
 & \textbf{Improv.}&\textbf{+2.24\%}&\textbf{+7.03\%}&\textbf{+5.04\%}& \textbf{+4.70\%}& \textbf{+2.20\%}&\textbf{+6.51\%}& & \textbf{+0.33\%}& \textbf{+5.36\%}&\textbf{+1.66\%}& \textbf{+4.21\%}& \textbf{+13.33\%}&\textbf{ +7.43\% }\\ 
                    \midrule
         \multirow{13}{*}{Clothing}&  HRLI@10 &   
0.0147&  0.9841&0.0341
&  0.9845&  0.5858& 0.9841
&  \multirow{13}{*}{ML-1M}& 0.4624& 0.7485&0.5283
&  0.6456& 0.7016& 0.8457\\ 
                    \cmidrule{2-8}
                    \cmidrule{10-15}
                    &  NDCG@5 &    0.0033&  0.0100&0.0103
&  0.0106& 0.0034& 0.0098
&  &0.1350& 0.1333&0.1064
&  0.1457& 0.0439& 0.0981\\ 
                    &  NDCG*@5&  0.0033&  0.0143&0.0104
& 0.0150&  0.0045& 
0.0139&  &0.1425& 0.1471&0.1116
& 0.1571& 0.0555& 0.1164\\
 & \textbf{Improv.}& +0.00\%& \textbf{+43.00\%}&\textbf{+0.97\%}& \textbf{+41.51\%}& \textbf{+32.35\%}& \textbf{+41.84\%}& & \textbf{+5.56\%}& \textbf{+10.35\%}&\textbf{+4.89\%}& \textbf{+7.82\%}& \textbf{+26.42\%}& \textbf{+18.65\% } \\ 
                    \cmidrule{2-8}
                    \cmidrule{10-15}
                    &  NDCG@10 &    0.0046&   0.0131&0.0119
& 0.0144&   0.0078& 0.0131&  &0.1646&  0.1627&0.1286
& 0.1733& 0.0707& 0.1238\\ 
                    &  NDCG*@10& 0.0047&  0.0173&0.0120&  0.0187&  0.0086& 
 0.0169&  & 0.1712& 0.1752&0.1335
& 0.1841& 0.0817& 0.1408\\
 & \textbf{Improv.}& \textbf{+2.17\%}& \textbf{+32.06\%}&\textbf{+0.84\%}& \textbf{+29.86\%}& \textbf{+10.25\%}& \textbf{+29.01\%}& &\textbf{+4.01\%}& \textbf{+7.68\%}&\textbf{+3.81\%}& \textbf{+6.23\%}& \textbf{+15.56\%}&\textbf{+13.73\%}\\ 
                    \cmidrule{2-8}
                    \cmidrule{10-15}
                    &  Hit@5 &   0.0053&     0.0184&0.0143
& 0.0194&   0.0073& 0.0182
&  &0.1974& 0.1972&0.1570& 0.2141& 0.0828& 0.1560\\ 
                    &  Hit*@5& 0.0053&  0.0206&0.0146
& 0.0221&  0.0094& 
 0.0204&  &0.2073& 0.2124&0.1644
& 0.2280& 0.0955& 0.1719\\
 & \textbf{Improv.}& +0.00\%& \textbf{+11.96\%}&\textbf{+2.10\%}&\textbf{+13.92\%}& \textbf{+28.76\%}& \textbf{ +12.09\%}& & \textbf{+5.02\%}& \textbf{+7.71\%}&\textbf{+4.71\%}& \textbf{+6.49\%}& \textbf{+15.34\%}&\textbf{ +10.19\%}\\ 
                    \cmidrule{2-8}
                    \cmidrule{10-15}
                    &  Hit@10 &  0.0095&    0.0282&0.0194
& 0.0312& 0.0211& 0.0284&  &0.2894& 0.2884&0.2258
& 0.2993& 0.1664& 0.2359\\ 
                    &  Hit*@10&  0.0095&   0.0299&0.0195&  0.0336&  0.0220&  0.0297&  &0.2969& 0.2998&0.2325& 0.3119& 0.1772& 0.2478\\ 

 & \textbf{Improv.}& +0.00\%&\textbf{+6.03\%}&\textbf{+0.52\%}& \textbf{+7.69\%}&\textbf{+4.26\%}&\textbf{ +4.58\%}& &\textbf{+2.59\%}& \textbf{+3.95\%}&\textbf{+2.97\%}& \textbf{+4.21\%}& \textbf{+6.49\%}& \textbf{+5.04\%}\\
 
        \bottomrule
    \end{tabular}
    }
    \caption{NDCG* and Hit* denote results after assigning negative infinity relevance score to last item. Performance improvement is denoted in bold.}
    \label{hrli_table}
\end{table*}
We evaluate HRLI@10 across six established sequential recommendation models  \cite{gru4rec, SASRec, stamp, lightsans, core, FEARec}, ranging from prominent baseline models to recent high-performing models, with the results presented in Table \ref{hrli_table}. 
Experiments are conducted on widely experimented datasets: three Amazon\cite{amazon} datasets--Beauty, Sports, and Clothing--and MovieLens-1M\cite{movielens} dataset. 
These datasets are selected to illustrate the effects of varying sparsity levels: 
the Amazon datasets are sparser, while ML-1M is less so, leading to shorter sessions in Amazon datasets and longer ones in ML-1M. Further implementation details and dataset statistics are provided in Appendix \ref{Implementation details}.
The results in Table 1 reveal that HRLI@10 always exceeds Hit@10 across all datasets and models. This suggests the existence of strong recency bias in existing models, where the models frequently rank the last item in a session higher than the actual GT item. Specifically,  SASRec, LightSANs, and FEARec exhibit particularly high HRLI values, reaching up to 0.99 in sparse datasets. GRU4Rec and STAMP show comparatively lower HRLI values, indicating that they are less vulnerable to recency bias. 
\section{Impacts of high HRLI on Recommendation Performance}\label{section_3}
\cite{full_eval}  emphasizes the crucial importance of evaluating recommendation performance using the entire item set, rather than a limited sample of negative items, in order to correctly measure recommendation performance. 
Since then, it has been widely adopted in sequential recommendation models\cite{cl4sr, duorec, FEARec, diffurec} to evaluate their proposed methods with a full item set. 
In Section \ref{section_2}, we observed that HRLI always exceeds Hit Rate, indicating that models frequently rank the last item in a session higher than the actual GT item, 
in the widely adopted full item set evaluation.
In addition to this, another problem is derived that 
the high rank of the last item can lower evaluated recommendation performance results by displacing the GT item from the Top-K ranking, in that standard evaluation metrics like Hit Rate and NDCG reward models when the GT item appears within the Top-K ranked items, while NDCG rewards more if the rank of GT item is high. 
This can cause the GT item to miss the Top-K cutoff by a single rank due to the last item’s consistently high rank. Figure \ref{fig:column}  in Appendix \ref{appendix:figure}  provides a more intuitive depiction of this case.  \\
To demonstrate the frequency of this case, we conduct experiments where we remove the last item from the Top-K ranking by assigning it a relevance score of \textit{negative infinity}. The results are presented in Table \ref{hrli_table}  under the metrics Hit* and NDCG*, and significant performance gains across almost all datasets and models are observed. The magnitude of these improvements is proportional to the HRLI value: models with high HRLI, such as SASRec, LightSANs, and FEARec, show substantial performance gains, which at the maximum improvement of 43.02\% is observed in NDCG*@5 of SASRec experimented on Sports dataset. Meanwhile, models with lower HRLI, GRU4Rec and STAMP, exhibit only minor improvements. For instance, GRU4Rec on the Clothing dataset has the lowest HRLI value of 0.0147, corresponding to its little or even zero increase in performance, which are the only results that do not show performance increase. Additionally, performance gains in NDCG*@5 and Hit*@5 are greater than those in NDCG*@10 and Hit*@10 overall, which aligns with the fact that removing the last item from the Top-K has a greater effect when K is small, given the last item’s consistently high rank. While it is possible to measure and note HRLI*@10, that is, HRLI of Hit*@10, it would be redundant, as the value is always zero. 
Our experiments conclusively demonstrate that high recency bias, as measured by our proposed HRLI metric, leads to a decline in recommendation performance. Therefore, accurately measuring recency bias using HRLI is crucial, as it also directly impacts the evaluated performance of sequential recommendation systems, in proportion to its magnitude. 
\section{Conclusion}
We emphasize the detrimental effects of high recency bias, in that they result in user disengagement and failure to fully learn user preference from user interaction session. Therefore we suggest a novel metric, HRLI to measure this recency bias given its importance. Experiments show that high HRLI lowers evaluated performances of sequential recommendation models and that lowering HRLI leads to increased evaluated performances across all experimented models, thus reemphasizing the repeated importance of measuring recency bias using our proposed HRLI metric.

\begin{acks}
This work was supported by Institute of Information \& communications Technology Planning \& Evaluation (IITP) grant funded by the Korea government(MSIT) [NO.RS-2021-II211343, Artificial Intelligence Graduate School Program (Seoul National University)]
\end{acks}

\bibliographystyle{ACM-Reference-Format}
\bibliography{chat}


\begin{thebibliography}{15}


\ifx \showCODEN    \undefined \def \showCODEN     #1{\unskip}     \fi
\ifx \showDOI      \undefined \def \showDOI       #1{#1}\fi
\ifx \showISBNx    \undefined \def \showISBNx     #1{\unskip}     \fi
\ifx \showISBNxiii \undefined \def \showISBNxiii  #1{\unskip}     \fi
\ifx \showISSN     \undefined \def \showISSN      #1{\unskip}     \fi
\ifx \showLCCN     \undefined \def \showLCCN      #1{\unskip}     \fi
\ifx \shownote     \undefined \def \shownote      #1{#1}          \fi
\ifx \showarticletitle \undefined \def \showarticletitle #1{#1}   \fi
\ifx \showURL      \undefined \def \showURL       {\relax}        \fi
\providecommand\bibfield[2]{#2}
\providecommand\bibinfo[2]{#2}
\providecommand\natexlab[1]{#1}
\providecommand\showeprint[2][]{arXiv:#2}

\bibitem[Chang et~al\mbox{.}(2022)]%
        {recency_dropout}
\bibfield{author}{\bibinfo{person}{Bo Chang}, \bibinfo{person}{Can Xu}, \bibinfo{person}{Matthieu L{\^e}}, \bibinfo{person}{Jingchen Feng}, \bibinfo{person}{Ya Le}, \bibinfo{person}{Sriraj Badam}, \bibinfo{person}{Ed Chi}, {and} \bibinfo{person}{Minmin Chen}.} \bibinfo{year}{2022}\natexlab{}.
\newblock \showarticletitle{Recency Dropout for Recurrent Recommender Systems}.
\newblock \bibinfo{journal}{\emph{arXiv preprint arXiv:2201.11016}} (\bibinfo{year}{2022}).
\newblock


\bibitem[Du et~al\mbox{.}(2023)]%
        {FEARec}
\bibfield{author}{\bibinfo{person}{Xinyu Du}, \bibinfo{person}{Huanhuan Yuan}, \bibinfo{person}{Pengpeng Zhao}, \bibinfo{person}{Jianfeng Qu}, \bibinfo{person}{Fuzhen Zhuang}, \bibinfo{person}{Guanfeng Liu}, \bibinfo{person}{Yanchi Liu}, {and} \bibinfo{person}{Victor~S Sheng}.} \bibinfo{year}{2023}\natexlab{}.
\newblock \showarticletitle{Frequency enhanced hybrid attention network for sequential recommendation}. In \bibinfo{booktitle}{\emph{Proceedings of the 46th International ACM SIGIR Conference on Research and Development in Information Retrieval}}. \bibinfo{pages}{78--88}.
\newblock


\bibitem[Fan et~al\mbox{.}(2021)]%
        {lightsans}
\bibfield{author}{\bibinfo{person}{Xinyan Fan}, \bibinfo{person}{Zheng Liu}, \bibinfo{person}{Jianxun Lian}, \bibinfo{person}{Wayne~Xin Zhao}, \bibinfo{person}{Xing Xie}, {and} \bibinfo{person}{Ji-Rong Wen}.} \bibinfo{year}{2021}\natexlab{}.
\newblock \showarticletitle{Lighter and better: low-rank decomposed self-attention networks for next-item recommendation}. In \bibinfo{booktitle}{\emph{Proceedings of the 44th international ACM SIGIR conference on research and development in information retrieval}}. \bibinfo{pages}{1733--1737}.
\newblock


\bibitem[Harper and Konstan(2015)]%
        {movielens}
\bibfield{author}{\bibinfo{person}{F~Maxwell Harper} {and} \bibinfo{person}{Joseph~A Konstan}.} \bibinfo{year}{2015}\natexlab{}.
\newblock \showarticletitle{The movielens datasets: History and context}.
\newblock \bibinfo{journal}{\emph{Acm transactions on interactive intelligent systems (tiis)}} \bibinfo{volume}{5}, \bibinfo{number}{4} (\bibinfo{year}{2015}), \bibinfo{pages}{1--19}.
\newblock


\bibitem[Hidasi et~al\mbox{.}(2015)]%
        {gru4rec}
\bibfield{author}{\bibinfo{person}{Bal{\'a}zs Hidasi}, \bibinfo{person}{Alexandros Karatzoglou}, \bibinfo{person}{Linas Baltrunas}, {and} \bibinfo{person}{Domonkos Tikk}.} \bibinfo{year}{2015}\natexlab{}.
\newblock \showarticletitle{Session-based recommendations with recurrent neural networks}.
\newblock \bibinfo{journal}{\emph{arXiv preprint arXiv:1511.06939}} (\bibinfo{year}{2015}).
\newblock


\bibitem[Hou et~al\mbox{.}(2022)]%
        {core}
\bibfield{author}{\bibinfo{person}{Yupeng Hou}, \bibinfo{person}{Binbin Hu}, \bibinfo{person}{Zhiqiang Zhang}, {and} \bibinfo{person}{Wayne~Xin Zhao}.} \bibinfo{year}{2022}\natexlab{}.
\newblock \showarticletitle{Core: simple and effective session-based recommendation within consistent representation space}. In \bibinfo{booktitle}{\emph{Proceedings of the 45th international ACM SIGIR conference on research and development in information retrieval}}. \bibinfo{pages}{1796--1801}.
\newblock


\bibitem[Kang and McAuley(2018)]%
        {SASRec}
\bibfield{author}{\bibinfo{person}{Wang-Cheng Kang} {and} \bibinfo{person}{Julian McAuley}.} \bibinfo{year}{2018}\natexlab{}.
\newblock \showarticletitle{Self-attentive sequential recommendation}. In \bibinfo{booktitle}{\emph{2018 IEEE international conference on data mining (ICDM)}}. IEEE, \bibinfo{pages}{197--206}.
\newblock


\bibitem[Krichene and Rendle(2020)]%
        {full_eval}
\bibfield{author}{\bibinfo{person}{Walid Krichene} {and} \bibinfo{person}{Steffen Rendle}.} \bibinfo{year}{2020}\natexlab{}.
\newblock \showarticletitle{On sampled metrics for item recommendation}. In \bibinfo{booktitle}{\emph{Proceedings of the 26th ACM SIGKDD international conference on knowledge discovery \& data mining}}. \bibinfo{pages}{1748--1757}.
\newblock


\bibitem[Li et~al\mbox{.}(2023)]%
        {diffurec}
\bibfield{author}{\bibinfo{person}{Zihao Li}, \bibinfo{person}{Aixin Sun}, {and} \bibinfo{person}{Chenliang Li}.} \bibinfo{year}{2023}\natexlab{}.
\newblock \showarticletitle{Diffurec: A diffusion model for sequential recommendation}.
\newblock \bibinfo{journal}{\emph{ACM Transactions on Information Systems}} \bibinfo{volume}{42}, \bibinfo{number}{3} (\bibinfo{year}{2023}), \bibinfo{pages}{1--28}.
\newblock


\bibitem[Liu et~al\mbox{.}(2018)]%
        {stamp}
\bibfield{author}{\bibinfo{person}{Qiao Liu}, \bibinfo{person}{Yifu Zeng}, \bibinfo{person}{Refuoe Mokhosi}, {and} \bibinfo{person}{Haibin Zhang}.} \bibinfo{year}{2018}\natexlab{}.
\newblock \showarticletitle{STAMP: short-term attention/memory priority model for session-based recommendation}. In \bibinfo{booktitle}{\emph{Proceedings of the 24th ACM SIGKDD international conference on knowledge discovery \& data mining}}. \bibinfo{pages}{1831--1839}.
\newblock


\bibitem[McAuley et~al\mbox{.}(2015)]%
        {amazon}
\bibfield{author}{\bibinfo{person}{Julian McAuley}, \bibinfo{person}{Christopher Targett}, \bibinfo{person}{Qinfeng Shi}, {and} \bibinfo{person}{Anton Van Den~Hengel}.} \bibinfo{year}{2015}\natexlab{}.
\newblock \showarticletitle{Image-based recommendations on styles and substitutes}. In \bibinfo{booktitle}{\emph{Proceedings of the 38th international ACM SIGIR conference on research and development in information retrieval}}. \bibinfo{pages}{43--52}.
\newblock


\bibitem[Qiu et~al\mbox{.}(2022)]%
        {duorec}
\bibfield{author}{\bibinfo{person}{Ruihong Qiu}, \bibinfo{person}{Zi Huang}, \bibinfo{person}{Hongzhi Yin}, {and} \bibinfo{person}{Zijian Wang}.} \bibinfo{year}{2022}\natexlab{}.
\newblock \showarticletitle{Contrastive learning for representation degeneration problem in sequential recommendation}. In \bibinfo{booktitle}{\emph{Proceedings of the fifteenth ACM international conference on web search and data mining}}. \bibinfo{pages}{813--823}.
\newblock


\bibitem[Xie et~al\mbox{.}(2022)]%
        {cl4sr}
\bibfield{author}{\bibinfo{person}{Xu Xie}, \bibinfo{person}{Fei Sun}, \bibinfo{person}{Zhaoyang Liu}, \bibinfo{person}{Shiwen Wu}, \bibinfo{person}{Jinyang Gao}, \bibinfo{person}{Jiandong Zhang}, \bibinfo{person}{Bolin Ding}, {and} \bibinfo{person}{Bin Cui}.} \bibinfo{year}{2022}\natexlab{}.
\newblock \showarticletitle{Contrastive learning for sequential recommendation}. In \bibinfo{booktitle}{\emph{2022 IEEE 38th international conference on data engineering (ICDE)}}. IEEE, \bibinfo{pages}{1259--1273}.
\newblock


\bibitem[Zhao et~al\mbox{.}(2022)]%
        {recbole2.0}
\bibfield{author}{\bibinfo{person}{Wayne~Xin Zhao}, \bibinfo{person}{Yupeng Hou}, \bibinfo{person}{Xingyu Pan}, \bibinfo{person}{Chen Yang}, \bibinfo{person}{Zeyu Zhang}, \bibinfo{person}{Zihan Lin}, \bibinfo{person}{Jingsen Zhang}, \bibinfo{person}{Shuqing Bian}, \bibinfo{person}{Jiakai Tang}, \bibinfo{person}{Wenqi Sun}, \bibinfo{person}{Yushuo Chen}, \bibinfo{person}{Lanling Xu}, \bibinfo{person}{Gaowei Zhang}, \bibinfo{person}{Zhen Tian}, \bibinfo{person}{Changxin Tian}, \bibinfo{person}{Shanlei Mu}, \bibinfo{person}{Xinyan Fan}, \bibinfo{person}{Xu Chen}, {and} \bibinfo{person}{Ji{-}Rong Wen}.} \bibinfo{year}{2022}\natexlab{}.
\newblock \showarticletitle{RecBole 2.0: Towards a More Up-to-Date Recommendation Library}.
\newblock \bibinfo{journal}{\emph{arXiv preprint arXiv:2206.07351}} (\bibinfo{year}{2022}).
\newblock


\bibitem[Zhao et~al\mbox{.}(2021)]%
        {recbole}
\bibfield{author}{\bibinfo{person}{Wayne~Xin Zhao}, \bibinfo{person}{Shanlei Mu}, \bibinfo{person}{Yupeng Hou}, \bibinfo{person}{Zihan Lin}, \bibinfo{person}{Yushuo Chen}, \bibinfo{person}{Xingyu Pan}, \bibinfo{person}{Kaiyuan Li}, \bibinfo{person}{Yujie Lu}, \bibinfo{person}{Hui Wang}, \bibinfo{person}{Changxin Tian}, \bibinfo{person}{Yingqian Min}, \bibinfo{person}{Zhichao Feng}, \bibinfo{person}{Xinyan Fan}, \bibinfo{person}{Xu Chen}, \bibinfo{person}{Pengfei Wang}, \bibinfo{person}{Wendi Ji}, \bibinfo{person}{Yaliang Li}, \bibinfo{person}{Xiaoling Wang}, {and} \bibinfo{person}{Ji{-}Rong Wen}.} \bibinfo{year}{2021}\natexlab{}.
\newblock \showarticletitle{RecBole: Towards a Unified, Comprehensive and Efficient Framework for Recommendation Algorithms}. In \bibinfo{booktitle}{\emph{{CIKM}}}. \bibinfo{publisher}{{ACM}}, \bibinfo{pages}{4653--4664}.
\newblock


\end{thebibliography}

\appendix

\section{Additional Explanation for Section \ref{section_3} with Figure}
\label{appendix:figure}

\begin{figure}[ht!]
    \centering
    \includegraphics[width=0.75\linewidth]{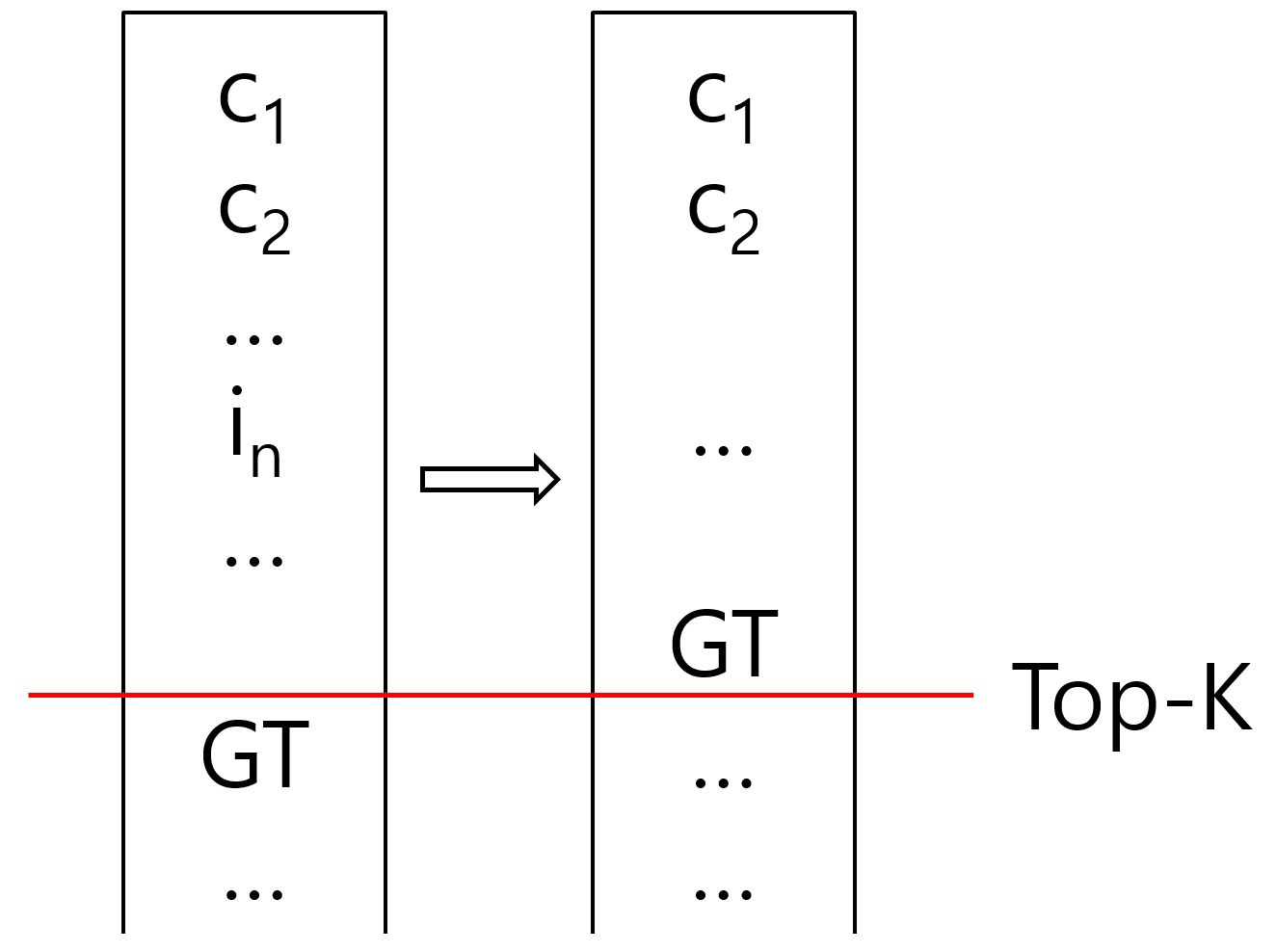}
    \caption{$c_i$ denotes candidate items, and $i_n$ denotes the last item in a session. }
    \label{fig:column}
\end{figure}
Figure \ref{fig:column} represents the case described in Section \ref{section_3}. 
Rankings are portrayed as bars in this figure. The left bar of the figure denotes the original status where the GT item has not been displaced from the Top-K ranking yet. The last item in the session, $i_n$, is taking place in the Top-K ranking, and the GT item is missing the Top-K cutoff by one rank. The right bar of the figure denotes after assigning a relevance score of negative infinity to $i_n$. Naturally, the rank of $i_n$ moves to the bottom of the ranking. Now that $i_n$ is displaced from the Top-K ranking, GT succeeds in passing the Top-K cutoff. The performance improvements reported in Table \ref{hrli_table} are rooted in this situation.

\section{Implementation Details and Dataset Statistics}
\label{Implementation details}
All our experiments are conducted on a popular recommendation framework RecBole\cite{recbole, recbole2.0}, with PyTorch. Seed is fixed to remove randomness and fully reproducible results. All models are experimented on hyper-parameter settings reported in their papers, according to each dataset. Maximum length of a session is set to 50 in all datasets, following \cite{duorec}. All other settings go by the default settings implemented in RecBole.
Items and users with less than five interactions are removed\cite{SASRec, duorec}, as widely adopted in sequential recommendation models. 
Early stopping is applied at all models on Hit@10, that if the best Hit@10 score on the validation set has not changed for ten epochs, training is stopped. \\
Dataset statistics after preprocessing are provided below. 

\begin{table}[h!]
    \small
    \begin{tabular}{cccccc}
    \toprule
         Dataset&  \# Users&  \# Items&  \# Inters&  Avg. Length& Sparsity\\
         \midrule
         Beauty&  22,364&  12,102&  198,502&  8.87& 99.92\%\\
         Clothing&  39,388&  23,034&  278,677&  7.07& 99.96\%\\
         Sports&  35,599&  18,358&  296,337&  8.32& 99.95\%\\
 ML-1M& 6,041& 3,417& 999,611& 165.49&95.15\%\\
 \bottomrule
    \end{tabular}
    \caption{Avg. Length denotes average session length.}
    \label{tab:my_label}
\end{table}


\end{document}